\documentclass[10pt]{article}

\usepackage{graphicx}%
\usepackage{amsmath,amssymb,amsthm,upref,bm}%
\usepackage{labelfig}%
\usepackage{mathrsfs} 

\DeclareMathAlphabet{\varmathbb}{U}{pxsyb}{m}{n}

\newtheorem{theorem}{Theorem}%

\newcommand{\pd}[3][]{\mathchoice{\raise-0.5pt\hbox{$\partial$}%
\vphantom{\partial}_{\mkern-1.5mu#2}^{\mkern0.4mu#1}\mkern0.3mu}%
{\raise-0.5pt\hbox{$\partial$}%
\vphantom{\partial}_{\mkern-1.5mu#2}^{\mkern0.4mu#1}\mkern0.3mu}%
{\raise-0.5pt\hbox{$\scriptstyle\partial$}%
\vphantom{\partial}_{\mkern-1.7mu#2}^{\mkern0.1mu#1}\mkern0.1mu}%
{\raise-0.5pt\hbox{$\scriptscriptstyle\partial$}%
\vphantom{\partial}_{\mkern-1.7mu#2}^{\mkern0.1mu#1}\mkern0.1mu}#3}
 
\newcommand{\D}{\mathrm{d}\kern0.2pt}%
\newcommand{\ii}{\kern0.05em\mathrm{i}\kern0.05em}%
\renewcommand{\vec}[1]{\bm{#1}}%

\begin{document}

\baselineskip=4.4mm

\makeatletter

\title{\bf On the absence of water waves \\ trapped near a cliffed cape}

\author{Nikolay Kuznetsov}

\date{}

\maketitle

\vspace{-6mm}

\begin{center}
Laboratory for Mathematical Modelling of Wave Phenomena, \\ Institute for Problems
in Mechanical Engineering, Russian Academy of Sciences, \\ V.O., Bol'shoy pr. 61,
St. Petersburg 199178, Russian Federation \\ E-mail: nikolay.g.kuznetsov@gmail.com
\end{center}

\begin{abstract}
The water wave problem is considered for a class of semi-infinite domains each
having its shore shaped as a cliffed cape. In particular, the free surface of a
water domain is supposed to be an infinite sector whose vertex angle is greater than
$\pi$, whereas the water layer lying under the free surface is of constant depth.
Under these assumptions, it is shown that there are no trapped mode solutions of the
problem for all values of a non-dimensional spectral parameter; in other words, no
point eigenvalues are embedded in the problem's continuous spectrum.
\end{abstract}

\setcounter{equation}{0}

\section{Introduction}

Let a non-dissipative medium occupy an unbounded domain, where waves can propagate.
By trapped modes (they are also referred to as resonance modes or bound states) one
understands localised free oscillations of the medium that cannot radiate to
infinity. More precisely, a time-harmonic solution of a boundary value problem
describing waves is called a trapped mode provided this solution decays at infinity
so that the total energy of the wave motion is finite. We just mention a couple of
recent articles that show a substantial progress in this area of research during the
past two decades. The absence of trapped modes in locally perturbed open acoustical
waveguides was considered in \cite{H}, whereas \cite{P} provides a survey of
localised modes in acoustics and elasticity. Some relevant results concerning water
waves are described in \S~1.1.

In this paper, we consider the linearised water wave problem for a class of
semi-infinite domains not investigated earlier. Namely, each of water domains under
consideration has an infinite sector as the free surface and its vertex angle is
$\pi + 2 \alpha$ with some $\alpha > 0$. Hence the complementary angle is less than
$\pi$, and so the coastline is cape-shaped. It is also assumed that the water layer
under the free surface is of constant depth, say $h$. Apart from this layer, the
water domain can include some subdomains covered from above by the rigid underwater
surface of the cape. In this case, its boundary has the form of an overhanging cliff
on one or both sides of the cape. Moreover, underwater tunnels through the rigid
body of the cape that connect its opposite sides are admissible. This kind of
geometry essentially distinguishes from those for which the absence and existence of
trapped modes were studied earlier (see \S~1.1).

For described domains, it is shown that there are no trapped mode solutions for all
values of a non-dimensional spectral parameter in the Steklov boundary condition on
the free surface at its mean position. This result provides a basis for proving the
uniqueness of solution for the following two problems. The first one is the
radiation problem describing the generation of outgoing waves by time-harmonic
vibrations of a finite part of the cape's underwater boundary. The second problem
concerns scattering of time-harmonic waves incoming from infinity.

\subsection{Background}

In the framework of the linear theory of water waves, the question of
absence/existence of trapped modes has a long history. Its initial point is the
classical paper \cite{John2} published by John in 1950. In this article, he proved
the first results guaranteeing the absence of trapped modes near bounded immersed
obstacles. In the case of a fixed one, the result is valid for all frequencies
provided some geometric restriction (now referred to as John's condition) is imposed
on the obstacle. Also, this condition yields the result for a freely floating
obstacle, but only for sufficiently large frequencies. In the three-dimensional
case, John's restriction on the geometry is as follows: the whole obstacle is
confined within the finite number of vertical cylinders through the lines, where
surface-piercing bodies (at least one such body must be a part of the obstacle)
intersect the free surface of water; moreover, the bottom (when the depth is
finite) is horizontal outside of these cylinders.

During the second half of the 20th century the main effort was devoted to finding
various conditions on geometry of fixed bodies guaranteeing the absence of trapped
modes (see the summarising monograph \cite{LWW} by Kuznetsov, Maz'ya and Vainberg).
In particular, it was Maz'ya who found the first such condition for totally
submerged bodies (see his papers \cite{Maz'ya,Maz} and \S~2.2 in \cite{LWW}). The
proof that his geometrical condition yields the absence of trapped modes is based on
a certain identity (now referred to as Maz'ya's identity). The latter proved to be
useful when establishing uniqueness in the water wave problems for various
geometries of immersed bodies, both fixed and floating freely; see, for example
\cite{KaM,CR,KM0,KM1}. The first two of these papers deal with fixed bodies, whereas
the second two with freely floating ones including totally submerged. Maz'ya's work
in the linear theory of water waves is not restricted to the just cited papers; a
detailed description of his results in this area can be found in \cite{KV} as well
as further references.

The first example of trapped mode was constructed by M.~McIver \cite{MM} only in
1996; her example involves two fixed surface-piercing cylinders separated by a
nonzero spacing. Another example of a mode trapped by two fixed surface-piercing
cylinders was found by Kuznetsov and Motygin (see \cite{LWW}, \S\,4.2.2.3).
Subsequently, Kuznetsov \cite{NGK10} proved that the latter cylinders can be
considered as two immersed parts of a single body which freely floats in trapped
waves remaining motionless; his result was generalised in \cite{KM}, where a brief
review of related papers is given. On the other hand, restrictions on the frequency
interval, that guarantee the absence of modes trapped by a freely floating body,
were obtained in \cite{K} and \cite{KM1} for two- and three-dimensional bodies
respectively. In both cases, the body's geometry must satisfy some assumptions;
actually, they are the same for freely floating bodies as for fixed ones.

\begin{figure}[t]
\begin{center}
\SetLabels
 \L (0.7*1.01) $F$\\
 \L (0.45*1.01) $C$\\
 \L (0.6*0.6) $W_1^{(\alpha)}$\\
 \L (0.33*0.44) $W_c$\\
 \L (0.0*0.84) The part of\\
 \L (0.0*0.72) cape with\\
 \L (0.0*0.6) overhanging\\
 \L (0.0*0.48) cliff\\
 \L (0.56*0.2) Rigid bottom\\
\endSetLabels \leavevmode \strut\AffixLabels{\includegraphics[width=70mm]{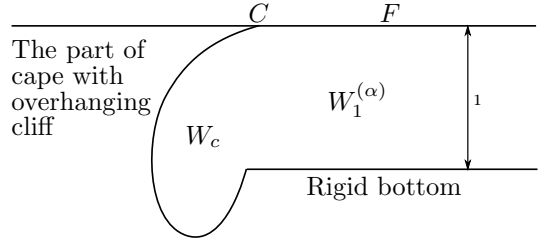}}
\end{center}
\vspace{-4mm} 
\caption{A sketch of the cross-section of the water domain by a plane
orthogonal to the coastline. The point, where the plane intersects the coastline, is
marked with $C$ above it. Various other cross-sections are marked as follows: $F$
denotes the free surface, whereas $W_c$ and $W_1^{(\alpha)}$ stand for the water
subdomains located under the overhanging cliff and the free surface respectively;
the latter has the constant depth equal to unity. }
\label{fig:1}
\vspace{-2mm} 
\end{figure}

\subsection{Statement of the problem; main result}

Let a Cartesian coordinate system $(\vec{x}, y)$ ($\vec{x}=(x_1,x_2)$ for the sake
of brevity) be such that the $y$-axis is directed upwards and the free surface at
rest, say $F$, lies in the plane $\{y=0\}$. Moreover, we assume that $F = \{
(\vec{x}, y) : \vec{x} \in F^{(\alpha)} , \, y=0 \}$, where
\[ F^{(\alpha)} = \{ \vec{x} \in \mathbb{R}^2 : \, x_2 > - |x_1| \tan \alpha  \} \quad
\mbox{and} \ \alpha \in (0 , \pi /2) .
\]
Thus, the free surface is a horizontal infinite sector whose vertex angle is $\pi +
2 \alpha$, whereas the mainland forms an infinite cape within the coast-line 
\[ C = \{ (\vec{x}, y) : \vec{x} \in \partial F^{(\alpha)} , \, y=0 \} . \]

It is convenient to formulate the problem of time-harmonic oscillations of water in
a dimensionless form. For this purpose we scale the coordinates $(x_1,x_2,y)$ so
that the depth of the water layer under $F$ is equal to unity. Thus, a water domain,
say $W$, is the union $W_1^{(\alpha)} \cup W_c$ (see Figure 1) of the layer
\begin{equation}
W_1^{(\alpha)} = \{ (\vec{x}, y) : \vec{x} \in F^{(\alpha)} , \, y \in (-1, 0) \}
\label{W}
\end{equation}
and a region $W_c$, which consists of one or more Lipschitz domains. This region is
such that $|\vec{x}|$ and $|y|$ are bounded for all its points. Moreover, $W_c$ is
adjoint to the rigid surface of the cape's overhanging cliff, that is, above the
level $y=-1$, each of the domains forming $W_c$ is confined between the vertical
half-planes, that go through the coast-line $C$ and form the dihedral angle $\pi - 2
\alpha$, but below the level $y=-1$ the geometry of $W_c$ is arbitrary. It is
admissible that $W_c$ is not simply connected; this takes place in the case of
underwater tunnel through the cape's body.

Following the procedure described in \cite{KM}, \S\,4, we consider the real-valued,
non-dimensional velocity potential $\phi$ as the problem's unknown. It is obtained
by using $\sqrt{h^3 g}$ as the scaling parameter; here $g > 0$ is the constant
acceleration due to gravity acting in the direction opposite to the $y$-axis and $h$
is the dimensional depth of the water layer under the free surface. Then the
potential $\phi$ describes a time-harmonic mode trapped in the domain $W$ provided
it is a finite energy solution of the homogeneous problem:
\begin{gather}
 \nabla^2 \phi = 0 \quad \mbox{in} \ W , \label{eq:1} \\
 \phi_y - \nu \phi = 0 \quad \mbox{on} \ F , \label{eq:2} \\
 \partial \phi / \partial n = 0 \quad \mbox{on} \ \partial W \setminus \bar F . 
 \label{eq:3}
\end{gather}
Here $\nabla = (\partial_{x_1}, \partial_{x_2}, \partial_{y})$ is the gradient
operator; $\nu = \omega^2 h / g$ is the non-dimensio\-nal spectral parameter related
to the radian frequency $\omega$ of water oscillations; $n$ denotes the normal on
$\partial W \setminus \bar F$ pointing outside of $W$.

It is convenient to understand the finiteness of energy in the sense that $\phi$
belongs to the usual Sobolev space $H^1 (W)$, in which the norm is as follows:
\[ \left[ \int_W |\nabla \phi|^2 \, \D \vec{x} \D y + \int_W |\phi|^2 \, \D \vec{x} \D y 
\right]^{1/2} .
\]
It is explained in \cite{NV} (see p. 3619) that this norm is equivalent to the
following one:
\[ \left[ \int_W |\nabla \phi|^2 \, \D \vec{x} \D y + \int_F |\phi|^2 \, \D \vec{x} 
\right]^{1/2} ,
\]
where the first (second) term in the square brackets is proportional to the kinetic
(potential) energy of the wave motion. Hence the fact that $\phi \in H^1 (W)$
expresses the finiteness of energy. Under this assumption problem
\eqref{eq:1}--\eqref{eq:3} is usually formulated in the weak sense:
\begin{equation}
 \int_{W} \nabla \phi \nabla \psi \,\D\vec{x}\, \D{}y = \nu
 \int_{F} \phi \, \psi \, \D\vec{x} .
\label{eq:intid}
\end{equation}
It is sufficient to require that this integral identity holds for all smooth
functions $\psi$ compactly supported in $\overline W$.

It is well known (see, for example, the book \cite{GT}, ch. 8), that the Laplace
equation holds in the classical sense provided $\phi \in H^1 (W)$ satisfies
\eqref{eq:intid}. The same concerns the following boundary conditions:
\begin{equation}
\phi_y - \nu \phi = 0 \quad \mbox{on} \ F \quad \mbox{and} \quad \phi_y = 0 \quad
\mbox{on} \ \partial W \cap \{y=-1\} .
\label{eq:-1}
\end{equation}
Now we are in a position to formulate the main result.

\begin{theorem}
Let $W = W_1^{(\alpha)} \cup W_c$ be the water domain, where $W_1^{(\alpha)}$ is
defined by formula \eqref{W} with $\alpha \in (0 , \pi /2)$ and $W_c$ is described
after that formula, and let $\nu$ be a positive number. If $\phi \in H^1 (W)$
satisfies problem \eqref{eq:1}--\eqref{eq:3} in the weak sense presented by identity
\eqref{eq:intid}, then $\phi$ vanishes identically in $W$.
\end{theorem}

This theorem means that there are no trapped water waves near a cliffed cape.


\section{Proof of Theorem 1}

The basic idea of the proof is to combine the classical method of John \cite{John2}
(see also \cite{LWW}, \S\,3.2.1.2) and the recent Rellich type theorem for the
Helmholtz equation obtained in \cite{BFHT}. Its version required for our purpose is
as follows.

\begin{theorem}
Let $k$ be a positive number. If $w \in L^2 (F^{(\alpha)})$, where $\alpha \in (0 ,
\pi /2)$, satisfies the Helmholtz equation $\nabla^2_{\vec{x}} w + k^2 w = 0$ in
$F^{(\alpha)}$ in the distributional sense, then $w$ vanishes identically; by
$\nabla_{\vec{x}}$ the gradient operator in $\mathbb{R}^2$ is denoted.
\end{theorem}

To prove that $\phi \in H^1 (W)$ vanishes identically when it satisfies problem
\eqref{eq:1}--\eqref{eq:3} in the domain $W$, on which some geometric assumptions
are imposed, it is sufficient to derive an inequality that contradicts the
equipartition of energy equality, namely:
\begin{equation}
\int_W |\nabla \phi|^2 \, \D \vec{x} \D y = \nu \int_F |\phi|^2 \, \D \vec{x} \, .
\label{eq:equi}
\end{equation}
Indeed, this relation is a direct consequence of the integral identity
\eqref{eq:intid} with $\psi = \phi$.

To implement this approach John based his method on properties of the following
auxiliary function:
\begin{equation}
w (\vec{x}) = \int_{-1}^0 \phi (\vec{x}, y) \cosh k_0 (y+1) \, \D y ,
\label{eq:aux}
\end{equation}
where $k_0$ is the unique positive zero of the function $\nu - k \tanh k$. Geometric
restrictions imposed on $W$ in Theorem~1 yield that this function is defined for
all $\vec{x} \in F^{(\alpha)}$. In view of equation \eqref{eq:1}, we have that
\[ \nabla^2_{\vec{x}} w = - \int_{-1}^0 \phi_{yy} (\vec{x}, y) \cosh k_0 (y+1) \, 
\D y \quad \mbox{in} \ F^{(\alpha)} \, ,
\]
and so after integration by parts twice in the right-hand side we obtain that
\begin{equation}
\nabla^2_{\vec{x}} w + k^2_0 w = 0 \quad \mbox{in} \ F^{(\alpha)} . 
\label{eq:nabla}
\end{equation}
Of course, the boundary conditions \eqref{eq:-1} must be taken into account for
cancelling the integrated terms.

For applying Theorem 2 to $w$, it remains to show that this function belongs to
$L^2 (F^{(\alpha)})$, which is true because $\phi \in L^2 (W)$. Indeed, the Schwarz
inequality gives
\begin{eqnarray*}
| w (\vec{x}) |^2 \leq \left( \int_{-1}^0 |\phi (\vec{x}, y)|^2 \, \D y \right)
\left( \int_{-1}^0 \cosh^2 k_0 (y+1) \, \D y \right) \\ = (2 k_0)^{-1} [ k_0 + \sinh
k_0 \cosh k_0 ] \int_{-1}^0 |\phi (\vec{x}, y)|^2 \, \D y \, .
\end{eqnarray*}
Therefore, $\int_{F^{(\alpha)}} | w (\vec{x}) |^2 \, \D \vec{x}$ is finite being
estimated by $\int_{W_1^{(\alpha)}} | \phi (\vec{x}, y) |^2 \, \D \vec{x} \D y$
with some positive constant, and we conclude that
\[ \int_{-1}^0 \phi (\vec{x}, y) \cosh k_0 (y+1) \, \D y = 0 \quad \mbox{for all}
\ \vec{x} \in F^{(\alpha)} .
\]

Integrating by parts again, we obtain
\[ \phi (\vec{x}, 0) \sinh k_0 = \int_{-1}^0 \phi_y (\vec{x}, y) \sinh k_0 (y+1) \,
\D y \quad \mbox{for all} \ \vec{x} \in F^{(\alpha)} ,
\]
and the Schwarz inequality gives
\begin{eqnarray*}
| \phi (\vec{x}, 0) \sinh k_0 |^2 \leq \left( \int_{-1}^0 |\phi_y (\vec{x}, y)|^2 \,
\D y \right) \left( \int_{-1}^0 \sinh^2 k_0 (y+1) \, \D y \right) \\ = \frac{1}{2}
\left[ \frac{\sinh^2 k_0}{\nu} - 1 \right] \int_{-1}^0 |\phi_y (\vec{x}, y)|^2 \, \D
y \, ,
\end{eqnarray*}
where the definition of $k_0$ is applied. An immediate consequence of this is the
inequality
\begin{equation}
\nu \int_{F} | \phi (\vec{x}, 0) |^2 \, \D \vec{x} \leq \frac{1}{2}
\int_{W_1^{(\alpha)}} |\nabla \phi|^2 \, \D \vec{x} \D y \, ,
\label{eq:con_rem}
\end{equation}
which is obviously incompatible with \eqref{eq:equi}. The proof of Theorem 1 is
complete.

\section{Concluding remarks}

The absence of trapped modes has been established for all values of a
non-dimensional spectral parameter under some geometrical restrictions on the water
domain surrounding a cliffed cape. The first key point of the proof is the
equipartition of energy equality \eqref{eq:equi}, whereas the auxiliary function
\eqref{eq:aux} used for deriving an inequality incompatible with \eqref{eq:equi} is
the second key point. Finally, it is essential that the function \eqref{eq:aux}
vanishes identically being a solution of equation \eqref{eq:nabla} and this follows
from Theorem 2 concerning the Helmholtz equation.

Since the proof of Theorem 1 does not involve any property of $W_c$ other than the
fact that it lies partly between $W_1^{(\alpha)}$ and the rigid cliff and partly
below $W_1^{(\alpha)}$, it is admissible to place a finite number of totally
submerged bounded rigid bodies into the region $W_c$, in which case Theorem~1 is
still true. Indeed, the equipartition of energy equality \eqref{eq:equi} remains
valid in this case as well as the considerations leading to inequality
\eqref{eq:con_rem}.

Let us discuss how Theorem 1 can be extended to a full-fledged uniqueness theorem
in the case when $W = W_1^{(\alpha)}$; that is, the water domain is the sectorial
layer, whose depth is constant and the vertex angle is $\pi + 2 \alpha$. By $k_1,
k_2, \dots$ we denote the sequence of positive zeroes of the function $\nu + k \tan
k$; this sequence up to the factor $\ii$ coincides with that of zeroes of $\nu - k
\tanh k$ lying on the positive imaginary axis. It is known (see, for example,
\cite{LMI}, \S\,8.4.1) that the sequence $\{ \psi_n (y) \}_{n=0}^\infty$ with
\[ \psi_0 (y) = \cosh k_0 (y+1) , \quad \psi_n (y) = \cos k_n (y+1) , \ \ n=1,2,\dots,
\]
is an orthogonal and complete system in $L^2 (-1, 0)$. Therefore, a solution of
equation \eqref{eq:1}, that satisfies the boundary conditions \eqref{eq:-1}, can be
written as follows:
\[ \phi (\vec{x}, y) = \sum_{n=0}^\infty v_n (\vec{x}) \psi_n (y) \quad \mbox{in}
\ W_1^{(\alpha)} .
\]
It is clear that we have in $F^{(\alpha)}$:
\begin{equation}
\nabla^2_{\vec{x}} v_0 + k^2_0 v_0 = 0 \quad \mbox{and} \quad \nabla^2_{\vec{x}} v_n
- k^2_n v_n = 0 \  \mbox{for} \ n = 1, 2, \dots \, .
\label{eq:nablan}
\end{equation}

Assuming that the norm of $\phi$ in $L^2 \big( W_1^{(\alpha)} \big)$ is finite, we
see that
\[ \int_{W_1^{(\alpha)}} |\phi|^2 \, \D \vec{x} \D y = \sum_{n=0}^\infty N_n^2
\int_{F^{(\alpha)}} |v_n|^2 \, \D \vec{x} , 
\]
where each integral on the right-hand side is finite and
\[ N_0^2 = \int_{-1}^0 \cosh^2 k_0 (y+1) \, \D y = \frac{1}{2} \left( 1 + \frac{\sinh
2 k_0}{2 k_0} \right) ,
\]
whereas
\[ N_n^2 = \int_{-1}^0 \cos^2 k_n (y+1) \, \D y = \frac{1}{2} \left( 1 + \frac{\sin
2 k_n}{2 k_n} \right) , \quad n=1,2,\dots \, .
\]
Then Theorem 2 yields that $v_0$ vanishes identically in $F^{(\alpha)}$. In the case
of the modified Helmholtz equation valid for $v_n$, $n = 1, 2, \dots$, the
uniqueness theorem analogous to Theorem~2 can be obtained in the same way as in
\cite{BFHT} (the proof is even simpler for this equation). Thus, every $v_n$ with $n
\geq 1$ also vanishes in $F^{(\alpha)}$. Therefore, if $\phi$ belongs to $L^2 \big(
W_1^{(\alpha)} \big)$, then $\phi$ vanishes identically regardless which boundary
condition holds on the vertical part of the domain's boundary, that is, on $\partial
W_1^{(\alpha)} \setminus (F \cup \{y=-1\})$.

Of course, in various uniqueness theorems proved for water wave problems involving
bounded obstacles in a layer of constant depth, a solution is supposed to belong to
a wider class than the corresponding $L^2$ space. Hence the requirement that the
$L^2$-norm of a solution is finite must be changed to a certain radiation condition
complementing problem \eqref{eq:1}--\eqref{eq:3} and guaranteeing that it has a
trivial solution only. In particular, such a condition can be formulated similarly
to that used for the two-dimensional Helmholtz equation in $F^{(\alpha)}$; see, for
example, \cite{BLG}, \S\,1.4, and compare with \cite{John2}, p.~48.

\end{document}